\documentclass[preprint,12pt,authoryear]{elsarticle}

\usepackage{amssymb}
\usepackage{amsmath}

\usepackage{algorithm}
\usepackage{algorithmic}
\usepackage{amssymb}
\usepackage{pifont}
\usepackage{booktabs}
\usepackage{array}
\usepackage{xcolor}
\usepackage{url}
\usepackage{graphicx}

\usepackage[normalem]{ulem} 

\journal{Expert Systems with Applications}

\begin{document}

\begin{frontmatter}




\title{Customized Tourist Navigation in Urban Contexts Following General Trends and Individual Preferences} 

\author[aff1]{Andrea Zingoni}
\author[aff1]{Sediola Ruko}
\author[aff2]{Enrique Yeguas-Bolívar}
\author[aff1,aff2]{José Manuel Alcalde-Llergo\corref{cor1}}
\ead{jose.alcalde@unitus.it}

\cortext[cor1]{Corresponding author}

\affiliation[aff1]{
  organization={Department of Economics, Engineering, Business and Society, University of Tuscia},
  addressline={Via Santa Maria in Gradi 4},
  city={Viterbo},
  postcode={01100},
  country={Italy}
}

\affiliation[aff2]{
  organization={Department of Computer Science and Artificial Intelligence, University of Córdoba},
  city={Córdoba},
  postcode={14071},
  country={Spain}
}


\begin{abstract}
Navigating an unfamiliar city poses significant challenges, and tourists are among the groups more likely to experience them, particularly when attempting to locate a point of interest (POI). Various factors, such as language barriers or a lack of precise information, further complicate this issue by making it difficult for visitors to explore efficiently. To address these issues, we introduce the tool ``Personalized Assistant for Tourist Hints'' (PATH), which is based on a methodology for estimating personalized tourist routes. PATH computes optimal paths to specified locations guided by two tailored heuristics: tourist frequency and POI preference. This integration enables efficient and personalized navigation, guiding users toward attractions that best match their preferences. We evaluated our methodology through a case study in Viterbo, Italy, using a dataset comprising over 1,000 routes from 265 tourists who visited the city during different periods. To validate our approach, we tested it with 238 new tourists and collected expert assessments from 86 local specialists. User satisfaction was measured through a structured questionnaire, which revealed highly positive feedback regarding the relevance and quality of the recommended routes, {also when compared with existing routing applications and approaches proposed in the literature}. These results demonstrate PATH’s potential to enhance urban exploration by providing an adaptable and user-centric solution for tourist navigation.
\end{abstract}




\end{frontmatter}



\section{Introduction}

Tourism plays an important role in global economies, fostering cultural exchange, economic growth, and regional development. The increasing accessibility of international travel has led to a surge in tourist activity, particularly in urban destinations rich in historic and cultural landmarks. However, exploring an unfamiliar city presents significant challenges for visitors, particularly those who do not speak the local language or are unfamiliar with the city's layout. Tourists often rely on traditional navigation tools such as digital maps or travel guides, which primarily offer generic routing solutions focused on efficiency, such as shortest or fastest paths~\citep{Souffriau2010}. While these approaches facilitate movement, they do not consider the preferences and interests of individual tourists, leading to a lack of personalization in the travel experience~\citep{Tussyadiah2017}.

A major limitation of conventional navigation systems is their inability to leverage the collective knowledge of past visitors to enhance route recommendations, as highlighted in ~\citep{Meng2021}. Popular navigation platforms typically compute paths using road network data and, in some cases, real-time traffic conditions, but they do not incorporate past data about tourist movement patterns~\citep{Zhang2024}. Yet, tourists often follow implicit movement trends, visiting points of interest (POIs) in certain sequences based on cultural appeal, accessibility, or thematic relevance. Integrating these movement patterns into routing decisions could significantly enhance tourist satisfaction by providing more contextually relevant and engaging exploration experiences~\citep{Sousa2024}. {Moreover, the approaches proposed so far in the literature are very limited in number and are often tailored to specific scenarios (particular geographical areas~\citep{XXX1}, mandatory round trips~\citep{XXX2}, multi-day itineraries~\citep{XXX3}, or group tourism~\citep{XXX4}). Furthermore, they are frequently not adequately validated from the perspective of user satisfaction~\citep{XXX1, XXX2, XXX3, XXX5}}.

To address these limitations, we introduce ``Personalized Assistant for Tourist Hints" (PATH), a tool for personalized tourist route planning that optimizes navigation by combining past tourist movement patterns with individual preferences for points of interest. The approach is based on a customized version of Dijkstra’s algorithm, incorporating two domain-specific heuristics. The first heuristic, tourist frequency, prioritizes routes frequently taken by past visitors, leveraging collective movement trends to suggest paths that are more likely to provide an enjoyable experience. The second heuristic, POI preference, aligns route recommendations with user-defined interests, ensuring that tourists visit locations that match their personal preferences rather than generic attractions. Only after these two aspects, and with lower priority, the system favors speed, namely, the time required to complete the entire route. By integrating the two heuristics into the routing process, PATH enhances personalization and adaptability, allowing tourists to explore cities in a way that aligns with their interests while benefiting from insights derived from past visitors.

{This work presents a novel system-level integration for tourist navigation. It builds upon established shortest-path foundations by introducing meaningful, data-driven heuristics that enhance route personalization and practical applicability in real-world scenarios.} In detail, the main contributions of this study are: (i) a tourist navigation framework that incorporates past visitor movement data and user-defined preferences into the route planning process, based on a large dataset of over 1,000 recorded tourist routes; (ii) a customized version of Dijkstra’s algorithm that integrates two heuristics, tourist frequency and POI preference, to optimize travel experiences; (iii) a validation through user studies, where 30 tourists tested and evaluated the system and 15 local experts assessed its practical utility, that gives a piece of evidence of the utility and usability of the proposed tool.

Unlike conventional routing methods that focus solely on minimizing distance or travel time, PATH introduces a more context-aware and user-centric navigation model that dynamically adapts to tourists' interests and time series data on tourist behavior. By leveraging collective movement data and enhancing traditional shortest-path algorithms with domain-specific heuristics, the proposed approach offers a more engaging and intuitive exploration experience for visitors.

{During the testing phase, our approach outperformed the most promising solutions proposed in the literature, which tend to focus on some specific case solely, while failing in general (limiting the potential and the applicability), and which lack a proper measurement of user satisfaction. In addition, tourists also preferred it with respect to the most used routing platforms, such as Google Maps, to enjoy an unfamiliar city}.

The remainder of this paper is structured as follows: Section 2 reviews related work on tourist navigation and personalized routing. Section 3 describes the methodology, including the customized routing algorithm and heuristics. Section 4 details the experimental setup and evaluation process. Section 5 presents the results and discusses them. Section 6 concludes with future research directions.

\section{Related Work}\label{RelatedWork}

Route optimization is a fundamental problem in graph theory and has been extensively studied in the context of transportation, logistics, and navigation systems. The problem of shortest path route calculation is central to many applications, ranging from vehicular navigation to pedestrian routing and robotic path planning. Classical algorithms, such as {Dijkstra’s ~\citep{Dijkstra1959}} and A* search ~\citep{Xiong2021,Rachmawati2020}, have been widely adopted due to their ability to compute optimal paths efficiently. While these methods are traditionally based on minimizing distance or travel time, they do not focus on user-specific preferences, dynamic environmental conditions, or domain-specific constraints. In recent years, various enhancements have been introduced to improve shortest path computation, including preprocessing techniques that accelerate query times ~\citep{su16020660}, as well as real-time adaptations that adjust routes based on traffic, weather, or pedestrian congestion~\citep{Qin2023}. However, despite these improvements, most shortest path algorithms are not inherently designed to incorporate high-level contextual factors such as tourist behavior, interest-based routing, or thematic exploration.

Given its robustness and optimality, the previously mentioned Dijkstra’s algorithm has been widely customized to suit different domain-specific routing problems. Numerous modifications have been proposed to enhance its performance, particularly for large-scale road networks where computational efficiency is critical~\citep{Alameri2024}. Some adaptations employ bidirectional search, hierarchical graph partitioning, or landmark-based heuristics to reduce search space complexity. Other approaches introduce dynamic weighting schemes that adjust edge costs based on external factors such as road congestion, toll costs, or accessibility constraints~\citep{udhan2022}. When coming to the tourism domain, modifications to Dijkstra’s algorithm have been explored to integrate user preferences and contextual data. These enhancements often involve heuristic-based filtering techniques that prioritize routes based on POIs relevance, scenic value, or cultural significance~\citep{Pei2022}. However,  such an approach completely neglects single user preferences, tending to guide tourists toward the most trendy visited places that do not always match personal tastes. Conversely, other navigation systems have been developed that extend classical routing algorithms by dynamically updating path recommendations based on evolving user interactions, as in~\cite{Ahmad2019}. In this case, relying only on the previous user's behavior leverages collective movement patterns, thus offering limited recommendations and risking the filter bubble effect ~\citep{filterbubbles}.

Beyond routing-focused applications, various technological advancements have been leveraged in tourism to improve visitor experiences. Digital tools such as mobile applications, augmented reality (AR), and recommendation systems have been developed to assist tourists in discovering attractions, understanding cultural heritage, and optimizing trip planning~\citep{Han2017,Yin2021}. Concerning AR, advanced navigation tools link movement data and boost the usability and performance of navigation applications, allowing for enhanced tourist experiences by facilitating better decision-making regarding routes and attractions~\citep{Pratisto2022}.
Concerning recommendation systems, these often rely on data mining and user profiling to enhance engagement; however, there is a notable gap as they rarely integrate real-world movement data to inform navigation choices. The findings in the literature indicate that while many applications suggest recommended places based on popularity or social media trends, there is a lack of integration with historically recorded patterns of tourist mobility that could further refine routing decisions~\citep{ValenciaArias2022}.

The integration of tourism-oriented algorithms in research has led to significant advancements in personalized navigation and recommendation systems. Many studies have focused on combining POI recommendations with route planning, leveraging techniques from artificial intelligence and geospatial data analysis~\citep{Hou2022}. Machine learning models have been employed to predict tourist behaviors~\citep{Zhou2020}, clustering users based on their movement patterns, interests, and social media interactions~\citep{Djebali2023}.  Moreover, reinforcement learning techniques have been applied to path problems, optimizing path selection based on feedback loops that adapt recommendations according to user engagement~\citep{Kong2022}. Multi-objective optimization approaches have also been explored, aiming to balance constraints such as time availability, POI popularity, and travel cost~\citep{Damos2021}. While these methods offer improved adaptability and personalization, their reliance on large-scale data training and high computational demands often limits their real-time applicability, particularly in dynamic urban environments~\citep{Hpken2020}. {In contrast to these complex architectures, our proposed PATH framework is positioned as an application-driven innovation. This computationally lightweight system achieves personalized routing without the need for extensive real-time model training, bridging the gap between theoretical multi-objective models and practical, deployable urban navigation.} 

{Some computationally lighter and more readily deployable solutions have been proposed in the literature in recent years. However, they remain limited in number and are not able to address the problem in its general form, as they typically focus only on specific scenarios. In~\citep{XXX1}, the Modified Adaptive Large Neighborhood Search (MALNS) method is used against running the Lingo program, with the specific aim of designing a tour for a family in a large area of Thailand. In~\citep{XXX2}, instead, a genetic algorithm is enhanced to optimize round trips. However, no evidence is provided regarding the applicability of these approaches in less context-specific scenarios. In~\citep{XXX3}, the authors employ Monte Carlo simulated annealing to search for the best solutions in multi-day trips, but user preferences are neglected during the optimization process. In~\citep{XXX4}, this aspect is taken into account, and the authors rely on a knowledge-based version of the ant colony algorithm to reconcile the preferences of groups of tourists during group travel. The use of ant colony optimization is also found in~\citep{XXX5}, although combined with a dynamic approach aimed at increasing the diversity of possible paths and avoiding convergence to local minima. While this solution appears to yield promising results in optimizing transportation for tourism, it is not suitable for optimizing pedestrian routes.
In addition to being limited to highly targeted scenarios, all these works also suffer from a major drawback: none of them has been effectively evaluated from the perspective of user satisfaction. In fact, validation has been generally based on the optimization of algorithmic parameters~\citep{XXX3, XXX4}, comparisons with similar but outdated solutions~\citep{XXX1}, or assessments conducted by the authors themselves against alternative routes~\citep{XXX2, XXX5}}.

This study builds upon these foundations, presenting a methodology that attempts to overcome the limitations of the currently available solutions to provide tourists with personalized routes that match their interests. This is done by integrating time series data about tourists' trajectories with individual preferences. By leveraging collective movement trends and, at the same time, aligning route recommendations with user interests, a more adaptive and context-aware navigation experience is offered. The evaluation of the proposed methodology in a real-world scenario, which took place in the city of Viterbo, located in the central part of Italy, and involved both tourists and experts in the tourism sector from the chosen city, demonstrated its potential to enhance urban exploration in a meaningful and scalable manner.

\section{Methodology}\label{sec:methodology}

In our work, the problem of tourist route recommendation is modeled as a graph-based pathfinding task~\citep{TTmetroxraine}, where typical movement patterns of tourists in general and the specific POIs for each user are jointly exploited to generate personalized navigation routes. The objective is to identify paths through a city that not only minimize travel distance but also maximize the relevance of visited locations according to the user's individual interests.

\begin{figure}[h]
    \centering
    \includegraphics[width=0.95\textwidth]{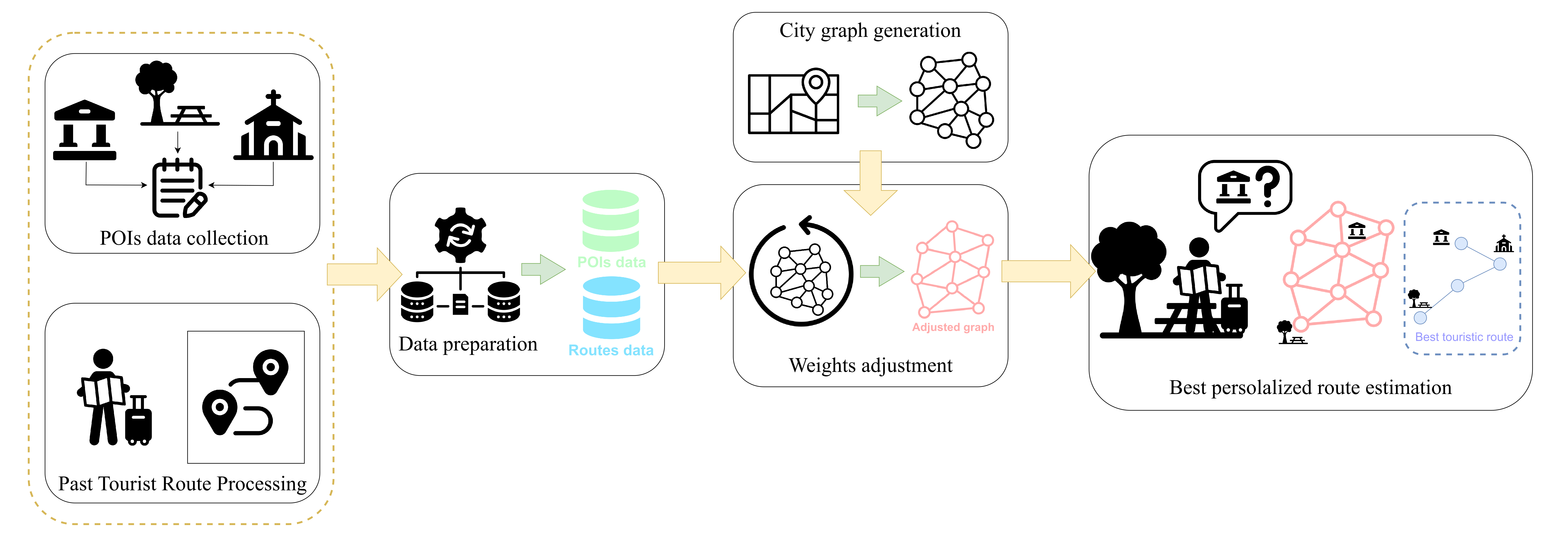}
    \vspace{-0.8em}
    \caption{Block diagram of the proposed methodology to generate personalized touristic routes.}
    \label{fig:overview}
\end{figure}

The methodology proceeds through several key stages. First, a dataset is built by collecting and processing information about specific POIs and general tourist routes. These data are used to construct a weighted graph, where nodes correspond to geographic locations and edges represent walkable paths. Such paths are initially weighted by physical distance, to include also this information in the final choice of the best paths, jointly with POIs and tourists' historical behavior.
To introduce personalization and contextual awareness, two ad-hoc heuristics are integrated into the graph:  frequency, which prioritizes frequently traversed paths from past data, and {POIs preference}, which aligns route suggestions with the specific interests expressed by each tourist individually. These heuristics dynamically adjust the edge weights to favor paths that are both among the most popular and match user profiles.
The final route computation is performed using a customized version of Dijkstra’s algorithm, which is modified to incorporate the heuristic-enhanced edge costs. As a result, the generated routes are capable of balancing time efficiency with popularity and personalized interests, offering tourists an enhanced exploration experience. An overview of the proposed methodology is depicted in Figure~\ref{fig:overview}, whereas, in the following subsections, each stage that comprises it is described in detail: data acquisition, graph construction, heuristic modeling, and, finally, personalized route computation.

\subsection{Data Collection}
To enable personalized tourist route generation, two distinct types of data are collected: (i) trajectories recorded from the movement of previous tourists and (ii) detailed information about POIs in the target city. These two pieces of information provide the basis for modeling navigation and enabling personalized path recommendations, based on both general information about typical tourists' behavior and individual preferences.

Concerning historical data about the typical movements of tourists, the trajectories of people who visited the city of Viterbo during two common vacation periods in Italy were collected, namely from the 22nd of June 2024 to the 8th of September 2024, and from the 7th of December 2024 to the 6th of January 2025. Tourists were asked to record their routes using Google Maps by saving the directions followed during their visits to various POIs. These routes were exported as GPX (GPS Exchange Format) files to ensure compatibility with the geographic data processing tools and libraries employed in this study. Each GPX file encodes a sequence of latitude–longitude coordinates representing the path taken by a tourist. These traces need to be preprocessed to reconstruct ordered trajectories and then mapped onto the city graph, as explained in \ref{sec:Preprocessing}.

Concerning POIs, instead, comprehensive information was gathered for each relevant location within the study area. Each POI was characterized by attributes such as geographic coordinates, name, descriptive metadata, plus visitor ratings and annual visitation statistics. {Data sources included official tourism databases, public travel platforms, and online review systems, {such as Google Maps and TripAdvisor, plus Travel365 and 10cose.it --which are popular touristic platforms in Italy. The POIs extracted from these sources has been revised by experts of tourism in Viterbo, to achieve a final list.}
Based on the above-mentioned data sources, all POIs were classified into one of five predefined categories:
\begin{itemize}
    \item \textit{Museums}: Includes art museums, historic exhibitions, cultural centers, and archaeological collections that showcase the artistic and historic heritage of the city.
    \item \textit{Parks}: Covers public green spaces such as city parks, botanical gardens, and recreational areas intended for leisure and outdoor enjoyment.
    \item \textit{Religious Sites}: Comprises churches, cathedrals, mosques, temples, and other places of worship with spiritual significance.
    \item \textit{Historic Sites}: Includes the main architectures having historic significance, such as ancient city walls with surviving gates and towers, castles, buildings housing seats of power over the course of centuries, etc.
    \item \textit{Other}: Groups all the noteworthy locations that do not fall into the aforementioned categories but are still significant to tourists, such as squares, sculptures, viewpoints, etc.    
\end{itemize}
It is worth noting that a POI can fall into more than one of the listed categories. In this case, it is classified into all of them.
In total, 42 POIs were identified and are considered in our analysis. Figure~\ref{fig:poi_distribution} summarizes the distribution of the 42 selected tourist POIs across the five main categories considered.

\begin{figure}[ht]
    \centering
    \includegraphics[width=0.6\textwidth]{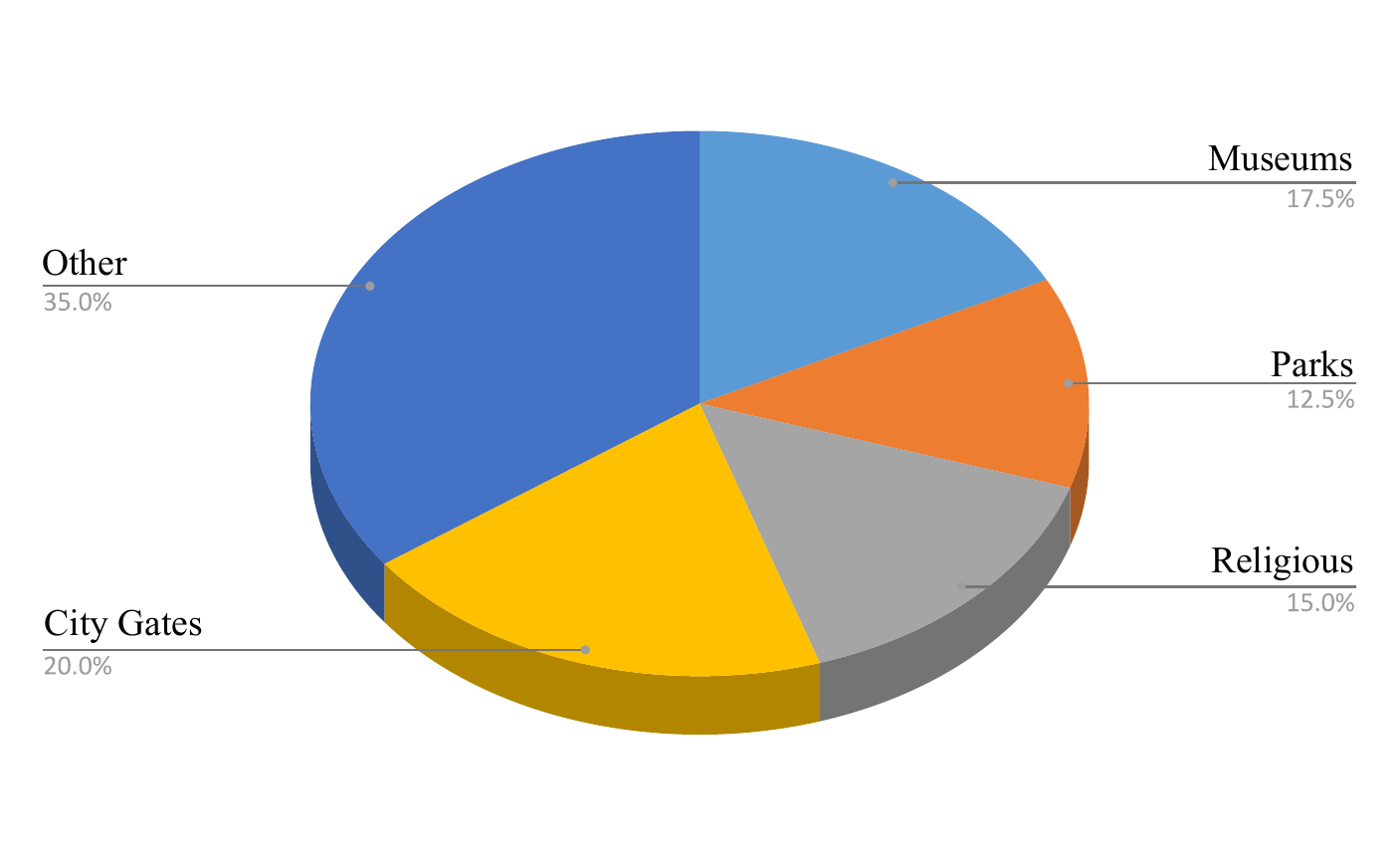}
    \caption{Distribution of the 42 identified tourist POIs across the five considered categories.}
    \label{fig:poi_distribution}
\end{figure}

\subsection{Routing Data Processing and Representation}\label{sec:Preprocessing}

Following the collection and preprocessing of tourist routes, the next step involves modeling the spatial structure of the city in a format suitable for route computation. To this end, the city is represented as a weighted graph, where locations and pathways are encoded using a node–edge structure that supports efficient pathfinding.
The graph is constructed by extracting the street network of the city using open-source geographic data provided by the OpenStreetMap platform, accessed through the \texttt{osmnx} library, as explained in~\cite{osmnx}. In this representation, nodes correspond to key geographic coordinates, typically street intersections or street ends, while edges represent the walkable paths that connect them. Each edge is initially weighted by an estimated travel time, derived from a database of walking speed averages, already present in the platform. Figure~\ref{fig:viterbo_graph} illustrates how the city of Viterbo is modeled as described above. Its graph is composed of 1858 nodes and 5392 edges.

\begin{figure}[h]
    \centering    \includegraphics[width=0.55\textwidth]{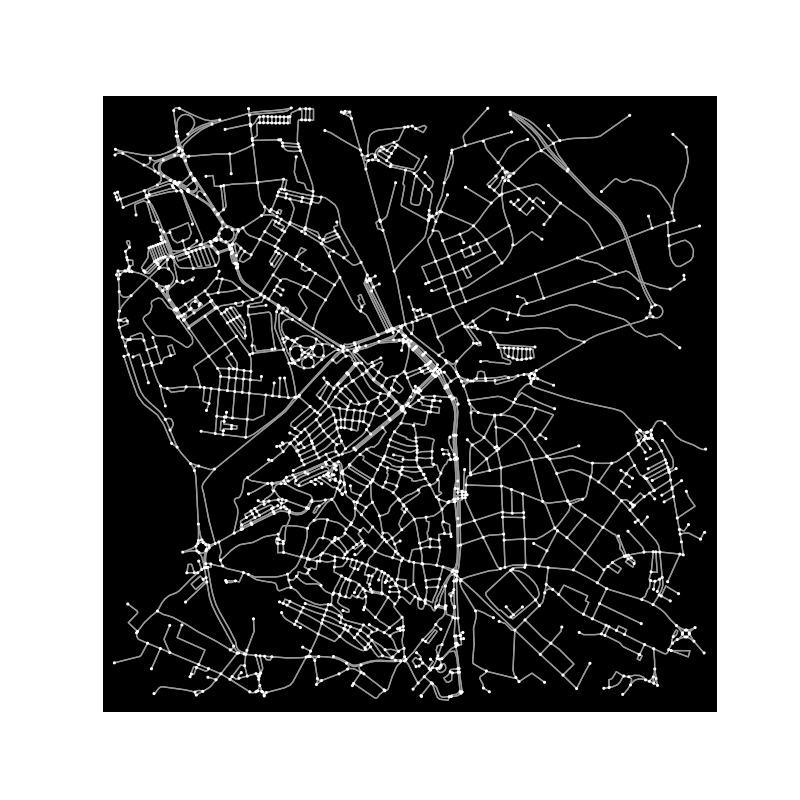}
    \caption{Graph-based spatial representation of the city of Viterbo.}
    \label{fig:viterbo_graph}
\end{figure}

The preprocessing of GPX paths basically consists of three operations. First, each trajectory is converted from a sequence of raw GPS coordinates into a sequence of node identifiers corresponding to the nodes from the underlying street network. Second, redundant nodes are removed. The map-matching process can sometimes generate consecutive duplicate nodes in the trajectory, particularly if multiple GPS points are projected onto the same network node (e.g., when a tourist stops at an intersection). These duplicates are removed to ensure a clean and efficient representation of the path, avoiding any misinterpretation in the frequency analysis. Third, any discontinuities in the route are resolved. Such gaps can appear due to signal loss in the original GPS trace (e.g., in narrow streets with tall buildings) or imperfections in the map-matching step, resulting in a sequence of nodes that is not directly connected in the graph. To ensure each route is a valid, continuous path, these gaps are bridged by computing and inserting the shortest path on the city graph between the two disconnected nodes. These corrections ensure that each tourist route is accurately and consistently represented as a sequence of valid walkable segments of the graph.

After that, in order to enhance the expressiveness of this model in touristic routes estimation and enable personalized recommendations, two domain-specific heuristics are integrated into the edge weighting mechanism. The first, referred to as tourist frequency ($\phi$), encodes how often a given path segment has been traversed by previous visitors of the city. This metric allows the system to prioritize commonly followed routes, capturing collective behavioral patterns. The second heuristic, tourist preferences ($\rho$), accounts for the individual interests of the current user by increasing the weight of connections associated with POI categories aligned with their profile.

The application of the two heuristics is made at different stages of the methodology. First, $\phi$ is applied during the construction of the network, enriching the static representation with information derived from previously recorded tourist routes. Specifically, each segment of the network is annotated with metadata reflecting how frequently it has been traversed, along with the distribution of POI categories visited along those paths. This process results in a structure that captures the movement patterns of past tourists and serves as the base for subsequent route computations.
In contrast, the heuristic $\rho$ is applied dynamically when computing a personalized route for a given user. Based on the current tourist’s interest profile, the influence of each POI category is taken into account to adjust the edge weights in a user-specific manner. As a result, the routing algorithm prioritizes paths that not only align with general popularity but also reflect the individual’s stated preferences.

Each tourist route \( r \in \mathcal{R} \) is also associated with a profile vector \( \hat{p}^{(r)} = \{p_i^{(r)}\}_{i \in P} \), where each component \( p_i^{(r)} \) represents the normalized proportion of POIs of category \( i \) visited along the route. The set \( P \) contains the predefined POI categories described earlier, and all vectors satisfy the normalization constraint \( \sum_{i \in P} p_i^{(r)} = 1 \). This representation enables each route to carry semantic information about tourist interests, which is then propagated into the network during the training process by incrementing the corresponding category-specific attributes on the edges traversed.

The Algorithm~\ref{alg:train_network} outlines the procedure used to train the graph with past data about tourists' behavior, encoding traversal frequency and semantic information related to POI categories across the visited edges. This algorithm processes a set of tourist trajectories that have already been preprocessed and mapped to the graph. For each edge traversed in a previous tourist route, it increments the frequency of use of this specific segment. In addition, it updates a set of category-specific weights using the POI category ratios associated with each route. These ratios represent the relative distribution of visited POI types and allow the network to encode not only how often a segment is used, but also the type of attractions typically associated with it.

\begin{algorithm}[ht]
\caption{Train Graph with Tourist Trajectories}
\label{alg:train_network}
\begin{algorithmic}[1]
\REQUIRE Network $G$, List of preprocessed routes $\mathcal{R}$, POI categories $P$
\ENSURE Enriched network $G_{\phi}$
\STATE $G_{\phi} \gets$ copy of $G$
\FOR{each route $r \in \mathcal{R}$}
    \FOR{each consecutive node pair $(u, v)$ in $r$}
        \IF{$(u,v)$ is an edge in $G_{\phi}$}
            \STATE $G_{\phi}[u,v].\text{count} \gets G_{\phi}[u,v].\text{count} + 1$
            \FOR{each category $p \in P$}
                \STATE $G_{\phi}[u,v].\text{category}[p] \gets G_{\phi}[u,v].\text{category}[p] + r.\text{poi\_ratio}[p]$
            \ENDFOR
        \ENDIF
    \ENDFOR
\ENDFOR
\RETURN $G_{\phi}$
\end{algorithmic}
\end{algorithm}

\subsection{Personalized Routing considering Tourist Preferences}

Once the graph has been adjusted with prior tourist frequency data ($\phi$), the next step is to generate personalized routes based on the individual preferences of each user. This is achieved by applying a second heuristic, denoted as $\rho$, which adjusts edge weights dynamically based on the different users' interest profiles.

Each tourist is associated with a profile vector \( \hat{t} = \{t_i\}_{i \in P} \), where \( t_i \in [0,1] \) represents the relative interest in POI category \( i \), and \( \sum_{i \in P} t_i = 1 \). This vector is initialized uniformly in the absence of prior information and progressively updated as the user visits different POIs during the route.

To incorporate this profile into the routing process, each edge of the enriched network contains accumulated interest values \( I_i \) for every POI category \( i \in P \), derived from past tourist routes. These values are normalized to compute an \textit{interest ratio} for each category:

\begin{equation}
r_i = \frac{I_i}{\sum_{j \in P} I_j}
\end{equation}

This ratio reflects the relative importance of POI category \( i \) along a specific edge. It enables the routing system to understand which types of attractions are most strongly associated with each edge of the graph.

To match this distribution with the interests of a specific user, the interest ratios are combined with the tourist’s profile vector using a weighted sum. The resulting value, called the \textit{Tourist Profile Factor} (TPF), is defined as:

\begin{equation}
TPF = \sum_{i \in P} (r_i \cdot t_i)
\end{equation}

{From a behavioral choice perspective, this linear combination follows a utility-based aggregation logic compatible with Multi-Attribute Utility Theory (MAUT). MAUT is a decision-making theory for evaluating and comparing alternatives based on several criteria or attributes~\citep{SONG_2024}. The TPF can be interpreted as an additive multi-attribute utility function in which the path segment represents the alternative, the category interest ratios $(r_i)$ represent the attribute-specific relevance scores, and the tourist's profile vector $(t_i)$ provides the subjective preference weights.} 

{Although the routing model incorporates a frequency heuristic~$(\phi)$ to capture collective popularity patterns, the TPF also introduces a personalization mechanism~$(\rho)$ that prioritizes semantic alignment with the individual user's interests. As a result, highly traversed paths are not systematically favored unless their associated POI composition is also relevant to the tourist profile. This balancing effect may contribute to reducing excessive concentration around generic tourist hotspots by encouraging routes with higher personal thematic relevance.}

{In summary,} TPF quantifies how well the semantic composition of a given edge aligns with the user’s preferences. Higher values indicate a stronger match between the edge and the tourist's interests.

To prioritize relevant segments, the original cost of each edge is penalized inversely proportional to the TPF. The adjusted cost \( L \), used in the shortest path computation, is calculated as:

\begin{equation}
L = L_0 \cdot (1 - \alpha \cdot TPF)
\end{equation}

Here, \( L_0 \) represents the original edge length or travel cost. To control the trade-off between path efficiency and thematic relevance, we introduce a tuning parameter, \( \alpha \in [0, 1] \), which weights the influence of the TPF. The final adjusted cost \( L \) is therefore calculated as:

\begin{equation}
L = L_0 \cdot (1 - \alpha \cdot TPF)
\end{equation}

This formulation allows the system to balance route directness with the user's interest profile. A value of \( \alpha = 0 \) would result in the shortest path, disregarding any preference, while a value closer to 1 would heavily prioritize routes aligned with the user's interests, at the expense of taking major detours.

The Algorithm~\ref{alg:generate_route} adjusts the weights of the trained graph \( G_\phi \) to reflect the current user's preferences, producing a personalized graph \( G_\rho \). For each edge, the total accumulated interest across all POI categories is computed. If no semantic information is available for a given edge, its original length is retained. Otherwise, the interest ratio \( r_i \) is calculated for each category and combined with the user's preference vector \( \hat{t} \) to compute the TPF. The edge length is then penalized by a factor proportional to \( 1 - TPF \), making edges that better match the user’s interests more likely to be included in the final route as they will be considered as shorter. The shortest path between the specified origin and destination is finally computed using Dijkstra’s algorithm over the adjusted graph.

\begin{algorithm}[ht]
\caption{Generate Personalized Route Based on Tourist Profile}
\label{alg:generate_route}
\begin{algorithmic}[1]
\REQUIRE Trained graph $G_\phi$, tourist profile vector $\hat{t}$, POI categories $P$, start node $s$, destination node $d$
\ENSURE Personalized route from $s$ to $d$
\STATE $G_\rho \gets$ copy of $G_\phi$
\FOR{each edge $(u,v)$ in $G_\rho$}
    \STATE $L_0 \gets G_\phi[u,v].\text{travel\_time}$
    \STATE $I_{\text{total}} \gets \sum_{i \in P} G_\phi[u,v].I_i$
    \IF{$I_{\text{total}} = 0$}
        \STATE $G_\rho[u,v].\text{weight} \gets L_0$
    \ELSE
        \STATE $TPF \gets 0$
        \FOR{each category $i \in P$}
            \STATE $r_i \gets G_\phi[u,v].I_i / I_{\text{total}}$
            \STATE $TPF \gets TPF + (r_i \cdot t_i)$
        \ENDFOR
        \STATE $G_\rho[u,v].\text{weight} \gets L_0 \cdot (1 - \alpha \cdot TPF)$
    \ENDIF
\ENDFOR
\STATE $route \gets$ Dijkstra($G_\rho, s, d$)
\RETURN $route$
\end{algorithmic}
\end{algorithm}

\subsection{Ethical Approval and Participant Consent}
All participants involved in the study{ (including those who generated the routes used for the graph adjustment and those recruited to test the application) }provided informed consent, ensuring they were fully aware of the research objectives and the handling of their data. The study complied with all ethical guidelines, including Organic Law 3/2018 on Personal Data Protection \citep{LOPD2018} and the ethical principles outlined in the Declaration of Helsinki~\citep{Helsinki2013}.

\section{Results and discussion}

This section presents an analysis of the results obtained through the proposed methodology, focusing on three key aspects: (i) visualization of collective tourist movement patterns, (ii) evaluation of personalized route generation, and (iii) user satisfaction assessment.

\subsection{Tourist Flow Visualization through Heat Maps}

The integration of tourist paths data into the city graph enables the generation of heat maps that visualize the most frequently traversed areas of the city. These maps provide valuable insights for urban planning and tourism management, offering a detailed understanding of how tourists navigate the city based on their interests. Such information can support strategic decisions regarding infrastructure improvements, crowd management, and cultural promotion.

Heat maps were generated for all the considered POI categories defined in the methodology. In Figure~\ref{fig:heatmaps}, three representative examples are presented. They refer to the categories {Museums}, {Historic Sites}, and {Other}, respectively.
The {Museums} category (Figure~\ref{fig:heatmaps}a) highlights classic tourist circuits, concentrating on heritage-rich areas and main exhibition sites. The {Historic Sites} map (Figure~\ref{fig:heatmaps}b) reflects movement along the city's ancient walls and the San Pellegrino neighborhood, a zone that preserved its medieval appearance. and key access points, offering insights into urban exploration routes linked to historic landmarks. Finally, the {Other Monuments} category (Figure~\ref{fig:heatmaps}c) reveals a more dispersed distribution, showcasing areas of secondary but still significant tourist interest.

The color gradient indicates the relative intensity of tourist flow, ranging from dark blue (low traffic) to red (high traffic). These visualizations exemplify how the methodology captures the semantic distribution of tourist activity, providing actionable data for local governance and enhancing the understanding of urban dynamics.

\begin{figure}[ht]
    \centering
    \includegraphics[width=0.95\linewidth]{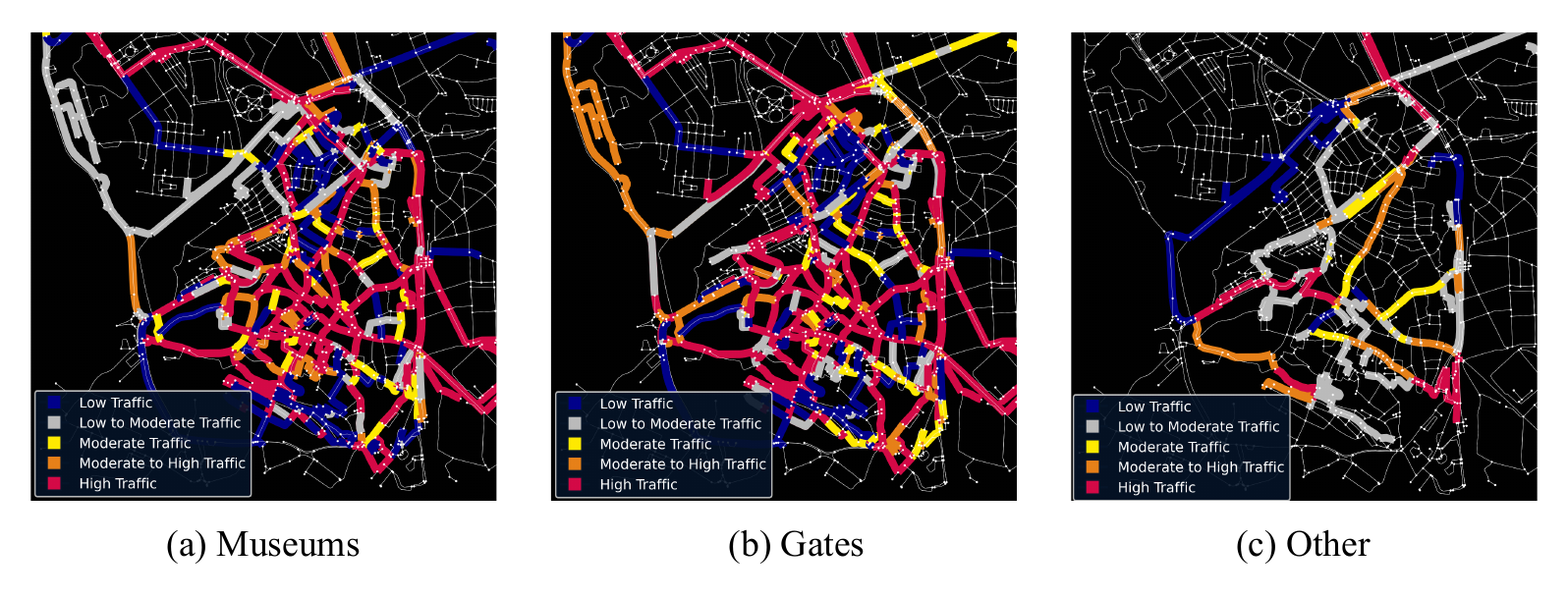}
    \caption{Heat maps representing tourist movement patterns in Viterbo, disaggregated by POI category: (a) Museums, (b) Historic Sites, and (c) Other Monuments. Colors indicate traversal frequency, with red denoting high-density areas and blue representing low-density zones.}
    \label{fig:heatmaps}
\end{figure}

\subsection{Generating personalized routes}

The personalized routing capabilities of the proposed approach were evaluated through a case study using the graph generated for the city of Viterbo. This graph was trained using a dataset of 1,023 tourist routes collected from 265 different visitors during summer and winter holidays, allowing the system to capture representative movement patterns and the semantic distribution of POI categories through the application of the $\phi$ heuristic. {Although the study is geographically limited to a single city, Viterbo serves as a representative test base. With approximately 60,000 inhabitants, considering 42 different POIs represents a highly dense mapping of its cultural heritage, characteristic of the historically rich pedestrian centers found in many European cities.} The collected travel trajectories may be made available upon reasonable request, in compliance with the constraints defined by the Ethics Committee that approved this study, and solely for research purposes. Access to the data can be requested by contacting the authors via the email addresses provided in this manuscript.

To illustrate the impact of PATH in generating personalized tourist routes, we must first define the setting for the tuning parameter $\alpha$. As introduced in Section 3.3, this parameter balances route efficiency against thematic POIs' interest. For this case study, and for the subsequent user evaluation, the parameter was empirically set to a value that allowed for detours of approximately 200 meters ($\alpha = 0.6$) from the shortest path in order to include relevant POIs. 

With this setting, a user profile vector was defined, containing the preference distribution about the POIs. An example of this vector is reported in Table~\ref{tab:simple_table}: in this case, the selected profile reflects a balanced interest across \textit{Museums}, \textit{Historic Sites}, and \textit{Other Monuments} categories, while assigning lower or null interest to \textit{Parks} and \textit{Religious Sites}.

\begin{table}[ht]
\caption{Example of a tourist profile vector used for personalized route generation.}
\label{tab:simple_table}
\centering
\begin{tabular}{lc}
\toprule
\textbf{POI Category} & \textbf{Preference Weight} \\ \midrule
Museums         & 0.3 \\
Parks           & 0.0 \\
Religious Sites & 0.1 \\
Historic Sites  & 0.3 \\
Other           & 0.3 \\
\bottomrule
\end{tabular}
\end{table}

Using this profile, the route between two of the principal gates of the ancient city of Viterbo, {Porta Romana} and {Porta Fiorentina}, was computed with the proposed methodology. Figure~\ref{fig:route_comparison}(a) offers a graphical representation of the obtained route, also highlighting the crossed POIs. It is worth noting that they belong to the most preferred categories, confirming the validity of the overall procedure.
Figure~\ref{fig:route_comparison}(b), instead, reports a comparison between the route calculated with the proposed methodology, inserting the $\phi$ and $\rho$ heuristics, and the route suggested by Google Maps, which optimizes only for travel time. 

\begin{figure}[h]
    \centering
    \includegraphics[width=0.9\textwidth]{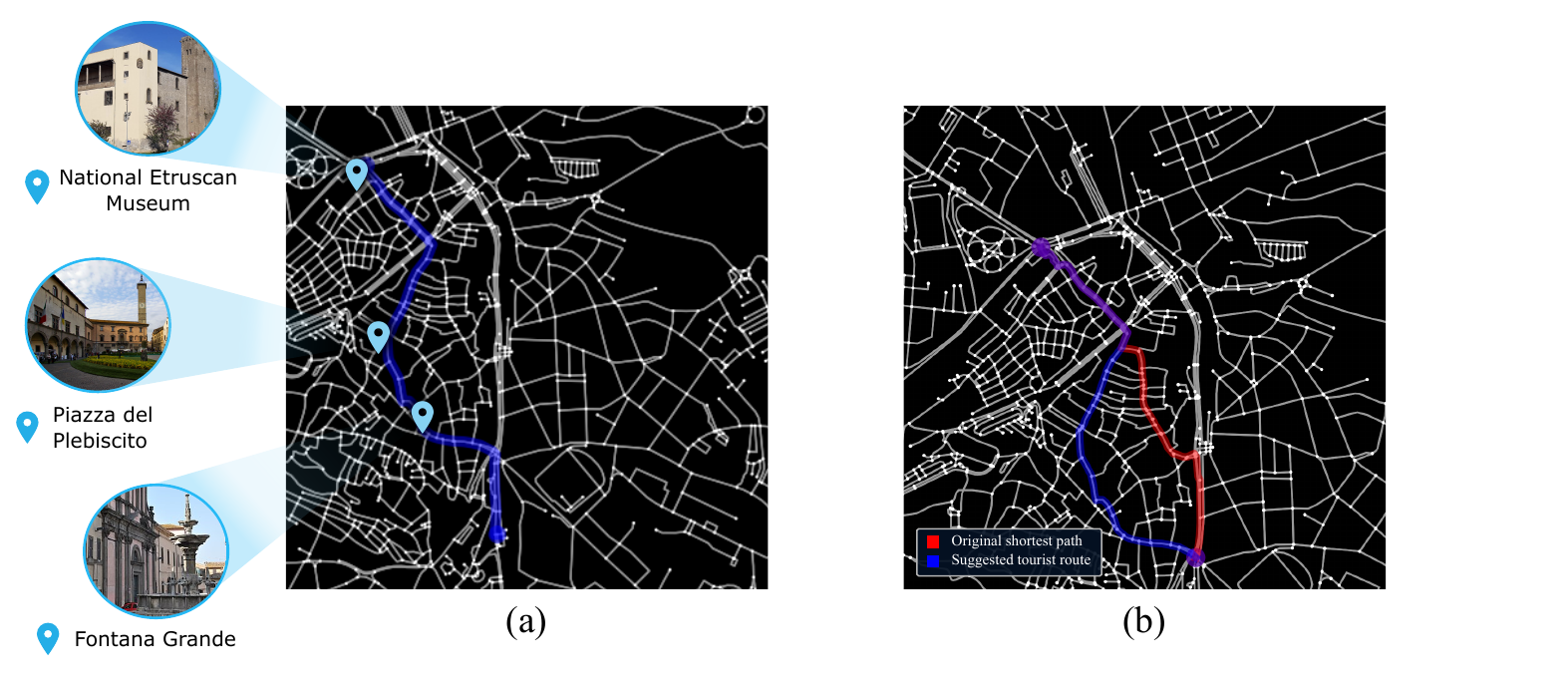}
    \caption{Comparison between the personalized route generated by the proposed method and the shortest path suggested by Google Maps. (a) Suggested route computed using the proposed model, highlighting the POIs traversed along the route. (b) Visual comparison of the two routes: the PATH route (blue), which incorporates tourist frequency and user preferences heuristics ($\phi$ and $\rho$), and the Google Maps route (red), which optimizes solely for travel time.}
    \label{fig:route_comparison}
\end{figure}

The comparison demonstrates that, while the proposed route may involve a slightly longer travel distance, it prioritizes paths that better align with the user's interests, providing a richer and more engaging tourist experience. This example underscores the potential of the methodology to offer context-aware navigation in urban environments, balancing efficiency with thematic relevance.

\subsection{User satisfaction evaluation}

{A noteworthy issue that emerged from our analysis of the existing scientific literature is that, to the best of our knowledge, no study reports an evaluation of user satisfaction in relation to the proposed approaches. Such an analysis is, however, essential for a comprehensive assessment of their actual usefulness and effectiveness for the intended target users, namely tourists. Thus, from November 2024 to February 2026, we conducted a study in Viterbo, Italy, aimed at assessing user satisfaction, both in absolute terms and through comparisons with alternative approaches. The study involved a total of 324 participants, divided into two groups: 268 tourists unfamiliar with the city and 56 local experts with extensive knowledge of its geography, POIs, and accessibility features, in order to provide a dual and complementary perspective. Participants tested the PATH solution and compared the routes generated by it with those provided by other systems. To ensure a comprehensive comparison, we considered approaches from both widely used routing applications and the most relevant and convincing methods available in the literature that could be reasonably replicated within our experimental setting. Among the routing applications, we considered widely used platforms such as Google Maps~\citep{googlemaps2024} and Apple Maps~\citep{applemaps2024}, as well as applications specifically designed for tourism, such as Sygic Travel Map~\citep{XXX6}, MyRoute-app~\citep{XXX7}, Roadtrippers~\citep{XXX8}, and Mindtrip~\citep{XXX9}. Among the approaches proposed in the literature, those presented in~\citep{XXX3} appeared to be the most promising candidates for comparison. To select suitable baselines, within the first group we initially excluded Apple Maps, as they are largely similar in purpose and functionality to Google Maps. We retained Google Maps as it is the most widely used navigation platform. We then analyzed tourism-oriented applications; however, these were designed for different use cases than ours. Specifically, Sygic and Mindtrip provide suggestions rather than routing functionalities, while MyRoute-app and Roadtrippers are tailored to longer trips and vehicle-based travel, rather than pedestrian urban exploration. Given these differences in scope, these applications were not included in the comparison. Within the second group, we encountered limitations related to the reproducibility of several approaches proposed in the literature. The method presented in~\citep{XXX3} was the only one that combined both reproducibility and relevance to the objectives of PATH, and was therefore selected for comparison.}

Feedback was collected through a structured questionnaire that addresses aspects such as satisfaction with the suggested route, perceived coverage of attractions, overall route preference, and comparison between PATH, Google Maps, and paper~\citep{XXX3} suggestions. The questionnaire comprises four items, summarized in Table~\ref{tab:questionnaire}. Different questions were addressed to tourists and experts, to capture both experiential and expert assessments; Table~\ref{tab:questionnaire} also reports the targeted category for each questionnaire item.

\begin{table}[ht]
\caption{Questionnaire for Participant Feedback. Tourist and expert respondents were asked different questions as indicated.}
\label{tab:questionnaire}
\centering
\resizebox{\textwidth}{!}{%
\begin{tabular}{c >{\raggedright\arraybackslash}p{10cm} c c}
\toprule
\textbf{No.} & \textbf{Questions} & \textbf{Tourist Question} & \textbf{Expert Question} \\ \midrule
1 & How satisfied are you with the attractions visited on the route suggested by PATH?  & \checkmark & \ding{55} \\
2 & Do you feel that PATH successfully included the points of interest matching your preferences? & \checkmark & \ding{55} \\
3 & Do you feel that the route suggested by PATH was worthy to be pursued, even if it slightly deviated from the shortest way? & \ding{55} & \checkmark \\
4 & Having used PATH, Google Maps, and the solution proposed in~\citep{XXX3}, which suggested provides a better tourist experience, in your opinion? & \checkmark & \checkmark \\
\bottomrule
\end{tabular}%
}
\end{table}

For the first three questions, responses were collected using a five-point Likert scale, where 1 indicates the lowest level of agreement or satisfaction and 5 indicates the highest. Figure~\ref{fig:questionnaire_results} summarizes the responses given by the participants.

\begin{figure}[h]
    \centering
    \includegraphics[width=0.9\linewidth]{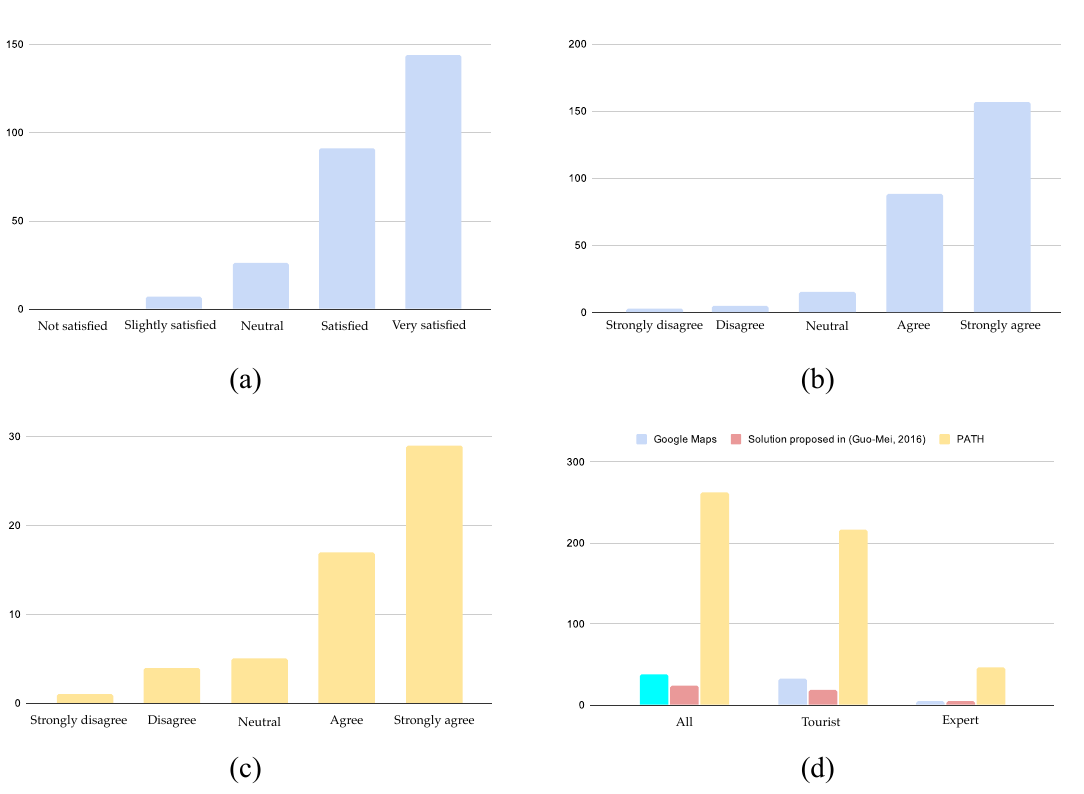}
    \caption{Summary of questionnaire responses from tourists and experts regarding the PATH system: (a) Answers to question 1, (b) Answers to question 2, (c) Answers to question 3, and (d) Answers to question 4.}
    \label{fig:questionnaire_results}
\end{figure}

When evaluating the satisfaction of tourists using PATH (Question 1), 54\% of them expressed the highest level (very satisfied) and around 34\% the second highest (satisfied). Less than 10\% expressed a neutral feeling, and only 3\% gave a negative response (and not the most negative). This result clearly provides the first evidence of the overall approval of the proposed methodology.

The approval is confirmed by the answers to Question 2, about the adequacy of the proposed system. Indeed, around 91.4\% of tourists rated the system with the two highest scores (4 and 5), indicating that only a few attractions may be missed in the suggested route.

Also, the feedback from local experts is in line with that of tourists, offering further support to the effectiveness and the global value of the proposed methodology. 
Question 3 gives indeed an indication of the practical value of the proposed system's deviations from the shortest path, which were governed by the \( \alpha \) parameter set for the study. Experts were asked if the resulting detours were worth pursuing. With 51.8\% rating the resulting path with a score of 5 and a further 30.4\% with a 4, the feedback strongly validates that PATH is capable of finding a meaningful balance between efficiency and cultural richness.

{Finally, when asked to choose between Google Maps, the approaches proposed in~\citep{XXX3} and PATH for an effective tourist experience (Question 4), 80.8\% of participants expressed a preference for the routes generated by PATH. In particular 80.5\% of tourists and even 82.1\% of local experts voted for PATH. Google Maps was preferred only by the 12.3\% of tourists and the 8.9\% of locals (with a total percentage of 11.7\%), whereas~\citep{XXX3} was voted by the 7.1\% of tourists and the 8.9\% of locals (with a total percentage of 7.4\%). These findings demonstrate the potential of the proposed methodology to enhance tourist navigation and satisfaction}.

\section{Conclusions and  Future Work}\label{sec:ConclusionsandFutureWork}

In this paper, we present a methodology to generate personalized tourist routes in urban environments by integrating typical tourist behavior and individual user preferences. The approach leverages a graph-based model enriched with two domain-specific heuristics: tourists' frequency, which captures collective route trends, and user preference, which allows dynamic personalization of routes based on the specific user's profiles. 

The proposed methodology was evaluated through a case study conducted in the city of Viterbo, starting from a dataset comprising more than 1,000 tourist routes collected from 265 visitors. Such a dataset allowed training a graph that, in turn, enabled the generation of heat maps that reveal frequent movement patterns across the city, specific to each category of monuments and attractions considered. This collection of heat maps can be considered as a first useful outcome of our work, since they provide valuable insights for urban planning and cultural heritage management. 

Additionally, the personalized routing capabilities of the proposed system were assessed through an ad-hoc questionnaire presented to tourists and local experts in the field of tourism, and focused on overall satisfaction and adequacy, as well as on a comparison with navigation applications that solely provide the shortest path, as Google Maps, {and approaches presented in the scientific literature that attempt, instead, to include tourist preferences in the suggested routes. The results demonstrated the system's ability to offer engaging and context-aware navigation experiences to the users, while balancing travel efficiency with thematic alignment to personal preferences. PATH was also preferred over Google Maps and outperformed similar approaches found in the literature by more than 80\% of users, confirming its value as a helpful tool for individuals who wish to explore an unfamiliar city}.

Future work will focus on extending the methodology to incorporate accessibility considerations, ensuring that the suggested routes are suitable for individuals with different kinds of disabilities. This line of work is inspired by previous research where navigation systems have been successfully adapted for inclusive purposes \citep{ruko1, pelorosso2024map4accessibility}. This enhancement aims to promote inclusive navigation experiences by integrating mobility constraints and accessibility requirements into the route recommendation process. {Although this first testing stage was limited to a single medium-sized city (Viterbo, with approximately 60,000 inhabitants and 42 evaluated POIs), scaling this methodology to larger metropolitan areas is a planned next step. In larger cities, where distances between POIs are significantly greater, the framework will need to be extended into a multi-modal routing system that incorporates public transportation networks. Nevertheless, the pedestrian-focused methodology proposed in this work could still be applied to the historical centers of those larger cities, as their spatial dimensions and POI densities are comparable to the urban topology evaluated in this study.}

\section*{CRediT authorship contribution statement}

\textbf{José Manuel Alcalde-Llergo}: Conceptualization, Methodology, Software, Formal analysis, Data Curation, Investigation, Writing – original draft, Visualization. 
\textbf{Sediola Ruko}: Conceptualization, Investigation, Resources, Data Curation, Writing – review \& editing
\textbf{Enrique Yeguas-Bolívar}: Validation, Writing – review \& editing, Supervision, Project administration.  
\textbf{Andrea Zingoni}: Validation, Resources, Investigation, Writing – review \& editing, Supervision, Project administration.

\bibliographystyle{elsarticle-harv} 
\bibliography{bibliography}

@article{osmnx,
author = {Boeing, Geoff},
year = {2017},
month = {07},
pages = {126-139},
title = {OSMNX: New Methods for Acquiring, Constructing, Analyzing, and Visualizing Complex Street Networks},
volume = {65},
journal = {Computers Environment and Urban Systems},
doi = {10.1016/j.compenvurbsys.2017.05.004}
}

@article{filterbubbles,
  title        = {Bias in algorithmic filtering and personalization},
  author       = {Bozdag, Engin},
  journal      = {Ethics and Information Technology},
  year         = {2013},
  volume       = {15},
  number       = {3},
  pages        = {209--227},
  month        = sep,
  publisher    = {Kluwer Academic Publishers},
  doi          = {10.1007/s10676-013-9321-6},
  address      = {Dordrecht, The Netherlands},
}

@INPROCEEDINGS{ruko1,
  author={Ruko, Sediola and Melloni, Daniele and Zingoni, Andrea and Pelorosso, Raffaele and Calabro, Giuseppe},
  booktitle={2024 IEEE International Conference on Metrology for eXtended Reality, Artificial Intelligence and Neural Engineering (MetroXRAINE)}, 
  title={Implementation of a Routing Application for People with Impairments, Improved and Evaluated Through Service Learning}, 
  year={2024},
  pages={547-552},
  doi={10.1109/MetroXRAINE62247.2024.10795951}}

@inbook{Souffriau2010,
  title = {Tourist Trip Planning Functionalities: State–of–the–Art and Future},
  ISBN = {9783642169854},
  ISSN = {1611-3349},
  DOI = {10.1007/978-3-642-16985-4\_46},
  booktitle = {Current Trends in Web Engineering},
  publisher = {Springer Berlin Heidelberg},
  author = {Souffriau,  Wouter and Vansteenwegen,  Pieter},
  year = {2010},
  pages = {474–485}
}

@article{Tussyadiah2017,
  title = {Embodiment of Wearable Augmented Reality Technology in Tourism Experiences},
  volume = {57},
  ISSN = {1552-6763},
  DOI = {10.1177/0047287517709090},
  number = {5},
  journal = {Journal of Travel Research},
  publisher = {SAGE Publications},
  author = {Tussyadiah,  Iis P. and Jung,  Timothy Hyungsoo and tom Dieck,  M. Claudia},
  year = {2017},
  month = may,
  pages = {597–611}
}

@article{Meng2021,
  title = {A Meaning-Aware Cultural Tourism Intelligent Navigation System Based on Anticipatory Calculation},
  volume = {11},
  ISSN = {1664-1078},
  DOI = {10.3389/fpsyg.2020.611383},
  journal = {Frontiers in Psychology},
  publisher = {Frontiers Media SA},
  author = {Meng,  Lei and Liu,  Yuan},
  year = {2021},
  month = jan 
}

@article{Zhang2024,
  title = {A Predictive Model Based on TripAdvisor Textual Reviews: Early Destination Recommendations for Travel Planning},
  volume = {14},
  ISSN = {2158-2440},
  DOI = {10.1177/21582440241246434},
  number = {2},
  journal = {Sage Open},
  publisher = {SAGE Publications},
  author = {Zhang,  Yating and Tan,  Hongbo and Jiao,  Qi and Lin,  Zhihao and Fan,  Zesen and Xu,  Dengming and Xiang,  Zheng and Law,  Rob and Zheng,  Tianxiang},
  year = {2024},
  month = apr 
}

@article{Xiong2021,
  title = {Application improvement of A* algorithm in intelligent vehicle trajectory planning},
  volume = {18},
  ISSN = {1551-0018},
  DOI = {10.3934/mbe.2021001},
  number = {1},
  journal = {Mathematical Biosciences and Engineering},
  publisher = {American Institute of Mathematical Sciences (AIMS)},
  author = {Xiong,  Xiaoyong and Min,  Haitao and Yu,  Yuanbin and Wang,  Pengyu},
  year = {2021},
  pages = {1–21}
}

@article{Sousa2024,
  title = {The Use of Artificial Intelligence Systems in Tourism and Hospitality: The Tourists’ Perspective},
  volume = {14},
  ISSN = {2076-3387},
  DOI = {10.3390/admsci14080165},
  number = {8},
  journal = {Administrative Sciences},
  publisher = {MDPI AG},
  author = {Sousa,  Ana Elisa and Cardoso,  Paula and Dias,  Francisco},
  year = {2024},
  month = aug,
  pages = {165}
}

@article{Alameri2024,
  title = {Optimizing Connections: Applied Shortest Path Algorithms for MANETs},
  volume = {141},
  ISSN = {1526-1506},
  DOI = {10.32604/cmes.2024.052107},
  number = {1},
  journal = {Computer Modeling in Engineering \& Sciences},
  publisher = {Tech Science Press},
  author = {Alameri,  Ibrahim and Komarkova,  Jitka and Al-Hadhrami,  Tawfik and Yahya,  Abdulsamad Ebrahim and Gharbi,  Atef},
  year = {2024},
  pages = {787–807}
}

@article{Rachmawati2020,
  title = {Analysis of Dijkstra’s Algorithm and A* Algorithm in Shortest Path Problem},
  volume = {1566},
  ISSN = {1742-6596},
  DOI = {10.1088/1742-6596/1566/1/012061},
  number = {1},
  journal = {Journal of Physics: Conference Series},
  publisher = {IOP Publishing},
  author = {Rachmawati,  Dian and Gustin,  Lysander},
  year = {2020},
  month = jun,
  pages = {012061}
}

@Article{su16020660,
AUTHOR = {Aburas, Hala and Shahrour, Isam and Giglio, Carlo},
TITLE = {Route Planning under Mobility Restrictions in the Palestinian Territories},
JOURNAL = {Sustainability},
VOLUME = {16},
YEAR = {2024},
NUMBER = {2},
ARTICLE-NUMBER = {660},
ISSN = {2071-1050},
DOI = {10.3390/su16020660}
}

@Article{Qin2023,
AUTHOR = {Qin, Zhicong and Pan, Younghwan},
TITLE = {Design of A Smart Tourism Management System through Multisource Data Visualization-Based Knowledge Discovery},
JOURNAL = {Electronics},
VOLUME = {12},
YEAR = {2023},
NUMBER = {3},
ARTICLE-NUMBER = {642},
ISSN = {2079-9292},
DOI = {10.3390/electronics12030642}
}

@misc{udhan2022,
      title={Vehicle Route Planning using Dynamically Weighted Dijkstra's Algorithm with Traffic Prediction}, 
      author={Piyush Udhan and Akhilesh Ganeshkar and Poobigan Murugesan and Abhishek Raj Permani and Sameep Sanjeeva and Parth Deshpande},
      year={2022},
      eprint={2205.15190},
      archivePrefix={arXiv},
      primaryClass={math.OC},
      url={https://arxiv.org/abs/2205.15190}, 
}

@article{Pei2022,
  title = {Optimization of tourism routes in Lushunkou District based on ArcGIS},
  volume = {17},
  ISSN = {1932-6203},
  DOI = {10.1371/journal.pone.0264526},
  number = {3},
  journal = {PLOS ONE},
  publisher = {Public Library of Science (PLoS)},
  author = {Pei,  Qian and Wang,  Li and Du,  Peng and Wang,  Zhaolan},
  editor = {Yu,  Vincent},
  year = {2022},
  month = mar,
  pages = {e0264526}
}

@ARTICLE{Ahmad2019,
  author={Ahmad, Shabir and Ullah, Israr and Mehmood, Faisal and Fayaz, Muhammad and Kim, DoHyeun},
  journal={IEEE Access}, 
  title={A Stochastic Approach Towards Travel Route Optimization and Recommendation Based on Users Constraints Using Markov Chain}, 
  year={2019},
  volume={7},
  number={},
  pages={90760-90776},
  doi={10.1109/ACCESS.2019.2926675}}

@article{Hou2022,
  title = {Research on Management Efficiency and Dynamic Relationship in Intelligent Management of Tourism Engineering Based on Industry 4.0},
  volume = {2022},
  ISSN = {1687-5265},
  DOI = {10.1155/2022/5831062},
  journal = {Computational Intelligence and Neuroscience},
  publisher = {Hindawi Limited},
  author = {Hou,  Tianchen},
  editor = {Chen,  Huihua},
  year = {2022},
  month = jan,
  pages = {1–9}
}

@article{Zhou2020,
  title = {Smart Tour Route Planning Algorithm Based on Naïve Bayes Interest Data Mining Machine Learning},
  volume = {9},
  ISSN = {2220-9964},
  DOI = {10.3390/ijgi9020112},
  number = {2},
  journal = {ISPRS International Journal of Geo-Information},
  publisher = {MDPI AG},
  author = {Zhou,  Xiao and Su,  Mingzhan and Liu,  Zhong and Hu,  Yu and Sun,  Bin and Feng,  Guanghui},
  year = {2020},
  month = feb,
  pages = {112}
}

@InProceedings{Kong2022,
author="Kong, Wei Kun
and Zheng, Shuyuan
and Nguyen, Minh Le
and Ma, Qiang",
editor="Strauss, Christine
and Cuzzocrea, Alfredo
and Kotsis, Gabriele
and Tjoa, A. Min
and Khalil, Ismail",
title="Diversity-Oriented Route Planning for Tourists",
booktitle="Database and Expert Systems Applications",
year="2022",
publisher="Springer International Publishing",
address="Cham",
pages="243--255"
}

@Article{Damos2021,
AUTHOR = {Damos, Mohamed A. and Zhu, Jun and Li, Weilian and Hassan, Abubakr and Khalifa, Elhadi},
TITLE = {A Novel Urban Tourism Path Planning Approach Based on a Multiobjective Genetic Algorithm},
JOURNAL = {ISPRS International Journal of Geo-Information},
VOLUME = {10},
YEAR = {2021},
NUMBER = {8},
ARTICLE-NUMBER = {530},
ISSN = {2220-9964},
DOI = {10.3390/ijgi10080530}
}

@article{Hpken2020,
  title = {Improving Tourist Arrival Prediction: A Big Data and Artificial Neural Network Approach},
  volume = {60},
  ISSN = {1552-6763},
  DOI = {10.1177/0047287520921244},
  number = {5},
  journal = {Journal of Travel Research},
  publisher = {SAGE Publications},
  author = {H\"{o}pken,  Wolfram and Eberle,  Tobias and Fuchs,  Matthias and Lexhagen,  Maria},
  year = {2020},
  month = jun,
  pages = {998–1017}
}

@article{Han2017,
  title = {User experience model for augmented reality applications in urban heritage tourism},
  volume = {13},
  ISSN = {1747-6631},
  DOI = {10.1080/1743873x.2016.1251931},
  number = {1},
  journal = {Journal of Heritage Tourism},
  publisher = {Informa UK Limited},
  author = {Han,  Dai-In and tom Dieck,  M. Claudia and Jung,  Timothy},
  year = {2017},
  month = feb,
  pages = {46–61}
}

@Article{Yin2021,
AUTHOR = {Yin, Celine Zhao Ying and Jung, Timothy and tom Dieck, M. Claudia and Lee, Maria Younghee},
TITLE = {Mobile Augmented Reality Heritage Applications: Meeting the Needs of Heritage Tourists},
JOURNAL = {Sustainability},
VOLUME = {13},
YEAR = {2021},
NUMBER = {5},
ARTICLE-NUMBER = {2523},
ISSN = {2071-1050},
DOI = {10.3390/su13052523}
}

@inproceedings{ValenciaArias2022,
  series = {ADVED 2022},
  title = {Technological Strategies For Knowledge Apprehension In Tourism-oriented Natural Environments: A Bibliometric Approach},
  DOI = {10.47696/adved.202229},
  booktitle = {Proceedings of ADVED 2022- 8th International Conference on Advances in Education},
  publisher = {International Organization Center of Academic Research},
  author = {Valencia-Arias,  Alejandro and Quiroz-Fabra,  Jefferson and Londoño,  Wilmer and Garcia-Pineda,  Vanessa and Rodríguez-Correa,  Paula and García Arango,  David},
  year = {2022},
  month = oct,
  collection = {ADVED 2022}
}

@article{Pratisto2022,
  title = {Immersive technologies for tourism: a systematic review},
  volume = {24},
  ISSN = {1943-4294},
  DOI = {10.1007/s40558-022-00228-7},
  number = {2},
  journal = {Information Technology \&; Tourism},
  publisher = {Springer Science and Business Media LLC},
  author = {Pratisto,  Eko Harry and Thompson,  Nik and Potdar,  Vidyasagar},
  year = {2022},
  month = jun,
  pages = {181–219}
}

@INPROCEEDINGS{TTmetroxraine,
  author={Ruko, Sediola and Alcalde-Llergo, José M. and Yeguas-Bolívar, Enrique and Zingoni, Andrea},
  booktitle={2024 IEEE International Conference on Metrology for eXtended Reality, Artificial Intelligence and Neural Engineering (MetroXRAINE)}, 
  title={Enhancing Tourist Experience via Automatic Personalized Route Suggestions}, 
  year={2024},
  volume={},
  number={},
  pages={698-703},
  doi={10.1109/MetroXRAINE62247.2024.10796573}}

@incollection{pelorosso2024map4accessibility,
  author    = {Pelorosso, Raffaele and Zingoni, Andrea and Ruko, Sediola and Calabrò, Giuseppe},
  title     = {{Map4Accessibility Project, An Inclusive and Participated Planning of Accessible Cities: Overview and First Results}},
  booktitle = {Innovation in Urban and Regional Planning},
  publisher = {Springer Nature Switzerland},
  year      = {2024},
  type      = {OriginalPaper | Chapter},
  url       = {https://www.springerprofessional.de/en/map4accessibility-project-an-inclusive-and-participated-planning/26778468}, 
}

@inbook{Djebali2023,
  title = {Hierarchical Clustering and Measure for Tourism Profiling},
  ISBN = {9783031251986},
  ISSN = {1611-3349},
  DOI = {10.1007/978-3-031-25198-6_12},
  booktitle = {Web and Big Data},
  publisher = {Springer Nature Switzerland},
  author = {Djebali,  Sonia and Gabot,  Quentin and Guerard,  Guillaume},
  year = {2023},
  pages = {158–165}
}

@misc{LOPD2018,
  author       = {{Government of Spain}},
  title        = {Organic Law on the Protection of Personal Data and Guarantee of Digital Rights},
  year         = {2018},
  note         = {Organic Law 3/2018, of December 5},
  howpublished = {\url{https://www.boe.es/eli/es/lo/2018/12/05/3/con}},
  urldate      = {2025-07-06}
}

@article{Helsinki2013,
  author  = {{World Medical Association}},
  title   = {World Medical Association Declaration of Helsinki: Ethical Principles for Medical Research Involving Human Subjects},
  journal = {JAMA},
  year    = {2013},
  volume  = {310},
  number  = {20},
  pages   = {2191--2194},
  doi     = {10.1001/jama.2013.281053}
}

@article{Dijkstra1959,
  title = {A note on two problems in connexion with graphs},
  volume = {1},
  ISSN = {0945-3245},
  url = {http://dx.doi.org/10.1007/BF01386390},
  DOI = {10.1007/bf01386390},
  number = {1},
  journal = {Numerische Mathematik},
  publisher = {Springer Science and Business Media LLC},
  author = {Dijkstra,  E. W.},
  year = {1959},
  month = dec,
  pages = {269–271}
}

@article{XXX1,
  title = {Modified ALNS Algorithm for a Processing Application of Family Tourist Route Planning: A Case Study of Buriram in Thailand},
  volume = {9},
  ISSN = {2079-3197},
  url = {http://dx.doi.org/10.3390/computation9020023},
  DOI = {10.3390/computation9020023},
  number = {2},
  journal = {Computation},
  publisher = {MDPI AG},
  author = {Khamsing,  Narisara and Chindaprasert,  Kantimarn and Pitakaso,  Rapeepan and Sirirak,  Worapot and Theeraviriya,  Chalermchat},
  year = {2021},
  month = feb,
  pages = {23}
}

@article{XXX2,
  title = {An Optimal Round-Trip Route Planning Method for Tourism Based on Improved Genetic Algorithm},
  volume = {2022},
  ISSN = {1687-5265},
  url = {http://dx.doi.org/10.1155/2022/7665874},
  DOI = {10.1155/2022/7665874},
  journal = {Computational Intelligence and Neuroscience},
  publisher = {Wiley},
  author = {Cao,  Sha},
  editor = {Kumar,  Vijay},
  year = {2022},
  month = aug,
  pages = {1–8}
}

@INPROCEEDINGS{XXX3,
  author={Guo-Mei, Hua},
  booktitle={2016 Eighth International Conference on Measuring Technology and Mechatronics Automation (ICMTMA)}, 
  title={Tourism Route Design and Optimization Based on Heuristic Algorithm}, 
  year={2016},
  volume={},
  number={},
  pages={449-452},
  doi={10.1109/ICMTMA.2016.113}}

@article{XXX4,
  title = {Tourism route optimization based on improved knowledge ant colony algorithm},
  volume = {8},
  ISSN = {2198-6053},
  url = {http://dx.doi.org/10.1007/s40747-021-00635-z},
  DOI = {10.1007/s40747-021-00635-z},
  number = {5},
  journal = {Complex \&; Intelligent Systems},
  publisher = {Springer Science and Business Media LLC},
  author = {Li,  Sidi and Luo,  Tianyu and Wang,  Ling and Xing,  Lining and Ren,  Teng},
  year = {2022},
  month = mar,
  pages = {3973–3988}
}

@inproceedings{XXX5,
  title = {Tourism Product Design and Development Based on Heuristic Algorithm},
  url = {http://dx.doi.org/10.1109/ICICTA.2015.268},
  DOI = {10.1109/icicta.2015.268},
  booktitle = {2015 8th International Conference on Intelligent Computation Technology and Automation (ICICTA)},
  publisher = {IEEE},
  author = {Hong,  Ye and Xiaochu,  Qu},
  year = {2015},
  month = jun,
  pages = {1067–1070}
}

@misc{XXX6,
  author       = {{Sygic}},
  title        = {Sygic Travel Maps},
  howpublished = {\url{https://www.sygic.com/travel}},
  year         = {2024},
  note         = {Accessed: 2026-03-15}
}

@misc{XXX7,
  author = {{MyRoute-app B.V.}},
  title = {MyRoute-app},
  year = {2024},
  howpublished = {\url{https://www.myrouteapp.com/en/}},
  note = {Accessed: 2026-03-15},
}

@misc{XXX8,
  author = {{Roadtrippers}},
  title = {Roadtrippers - Road Trip Planner},
  year = {2024},
  howpublished = {\url{https://roadtrippers.com}},
  note = {Accessed: 2026-03-15},
}

@misc{XXX9,
  author       = {{Mindtrip Inc.}},
  title        = {Mindtrip: AI-Powered Travel Planning and Booking Platform},
  howpublished = {\url{https://mindtrip.ai}},
  year         = {2024},
  note         = {Accessed: 2026-03-15}
}

@misc{googlemaps2024,
  author       = {{Google}},
  title        = {{Google Maps}},
  howpublished = {\url{https://www.google.com/maps}},
  year         = {2024},
  note         = {Accessed: 2026-03-15}
}

@misc{applemaps2024,
  author       = {{Apple}},
  title        = {{Apple Maps}},
  howpublished = {\url{https://maps.apple.com}},
  year         = {2024},
  note         = {Accessed: 2026-03-15}
}

@article{SONG_2024,
title = {Implementing multi-attribute utility theory in service recovery: An operational management perspective on online retailing},
journal = {Journal of Retailing and Consumer Services},
volume = {81},
pages = {103968},
year = {2024},
issn = {0969-6989},
doi = {https://doi.org/10.1016/j.jretconser.2024.103968},
url = {https://www.sciencedirect.com/science/article/pii/S0969698924002649},
author = {Yan Song and Yifan Xiu and Liping Zhou and Jingyuan Wang}
}






\end{document}